# Distinct Topological Surface States on the Two Terminations of MnBi$_4$Te$_7$

Xuefeng Wu[1#], Jiayu Li[1#], Xiao-Ming Ma[1#], Yu Zhang[1,2#], Yuntian Liu[1], Chun-Sheng Zhou[1], Jifeng Shao[1], Qiaoming Wang[1], Yu-Jie Hao[1], Yue Feng[1], Eike F. Schwier[3], Shiv Kumar[3], Hongyi Sun[1], Pengfei Liu[1], Kenya Shimada[3], Koji Miyamoto[3], Taichi Okuda[3], Kedong Wang[1], Maohai Xie[2], Chaoyu Chen[1], Qihang Liu[1,4*], Chang Liu[1*], and Yue Zhao[1*]

[1]Shenzhen Institute for Quantum Science and Engineering and Department of Physics, Southern University of Science and Technology, Shenzhen 518055, China
[2]Department of Physics, University of Hong Kong, Hong Kong, China
[3]Hiroshima Synchrotron Radiation Center, Hiroshima University, 2-313 Kagamiyama, Higashi-Hiroshima 739-0046, Japan
[4]Guangdong Provincial Key Laboratory for Computational Science and Material Design, Southern University of Science and Technology, Shenzhen 518055, China

---

[#] These authors contribute equally to this work.
[*] To whom correspondence should be addressed: zhaoy@sustech.edu.cn, liuc@sustech.edu.cn, liuqh@sustech.edu.cn.

**Abstract**

The recent discovered intrinsic magnetic topological insulator MnBi$_2$Te$_4$ have been met with unusual success in hosting emergent phenomena such as the quantum anomalous Hall effect and the axion insulator states. However, the surface-bulk correspondence of the Mn-Bi-Te family, composed by the superlattice-like MnBi$_2$Te$_4$/(Bi$_2$Te$_3$)$_n$ ($n$ = 0, 1, 2, 3…) layered structure, remains intriguing but elusive. Here, by using scanning tunneling microscopy (STM) and angle-resolved photoemission spectroscopy (ARPES) techniques, we unambiguously assign the two distinct surface states of MnBi$_4$Te$_7$ ($n$ = 1) to the quintuple-layer (QL) Bi$_2$Te$_3$ termination and the septuple-layer (SL) MnBi$_2$Te$_4$ termination, respectively. A comparison of the experimental observations with theoretical calculations reveals the diverging topological behaviors, especially the hybridization effect between magnetic and nonmagnetic layers, on the two terminations: a gap on the QL termination originating from the topological surface states of the QL hybridizing with the bands of the beneath SL, and a gapless Dirac-cone band structure on the SL termination with time-reversal symmetry. The quasi-particle interference patterns further confirm the topological nature of the surface states for both terminations, continuing far above the Fermi energy. The QL termination carries a spin-helical Dirac state with hexagonal warping, while at the SL termination, a strongly canted helical state from the surface lies between a pair of Rashba-split states from its neighboring layer. Our work elucidates an unprecedented hybridization effect between the building blocks of the topological surface states, and also reveals the termination-dependent time-reversal symmetry breaking in a magnetic topological insulator, rendering an ideal platform to realize the half-integer quantum Hall effect and relevant quantum phenomena.

# I. INTRODUCTION

The recent discovery of the intrinsic magnetic topological insulator $MnBi_2Te_4$ and its derivatives $MnBi_2Te_4/(Bi_2Te_3)_n$ ($n = 1, 2, …$), comprising alternating layers of $MnBi_2Te_4$ and nonmagnetic topological insulator (TI) $Bi_2Te_3$, has boosted exciting possibilities of producing exotic quantum phenomena such as the quantum anomalous Hall (QAH) effect and the topological magnetoelectric effect by engineering topology and magnetism at the atomic scale[1-6]. Specifically, $MnBi_2Te_4$, having antiferromagnetic (AFM) order and thus broken time-reversal symmetry, holds the potential for realizing both QAH insulator (or Chern insulator) and axion insulator[7-10]. When subjected to an external magnetic field of around 5-10 Tesla, few-layer $MnBi_2Te_4$ turns to a ferromagnetic (FM) Chern insulator[11]. On the other hand, zero Hall plateau is observed in even-layer $MnBi_2Te_4$ as an indicator for axion insulator[10]. However, recent ARPES measurements show a robust surface Dirac cone, indicating that time-reversal symmetry is preserved at the surface of $MnBi_2Te_4$[12-15]. Other members in van der Waals $MnBi_2Te_4/(Bi_2Te_3)_n$, with tunable interlayer magnetic exchange coupling[16-20], also exhibit controversially gap or gapless feature at different terminations[21-23], highlighting the complication of the interplay between magnetism and topology and the crucial role of defects and disorder in such a material system.

For $n \geq 1$, the hybrid structure creates distinct electronic structures on separate terminations due to the interplay of topology and magnetism at the interfaces of magnetic TI ($MnBi_2Te_4$) and non-magnetic TI ($Bi_2Te_3$) layers. Local probe approaches to the scattering processes[24-26] on different terminations of the van der Waals heterostructure are crucial to determine the spin configuration of the topological surface states and advance the theoretical understanding of the heterostructure engineering. Moreover, despite the evidence of the band structure in MnBi$_2$Te$_4$/(Bi$_2$Te$_3$)$_n$ series from ARPES and transport studies, so far there has been little experimental work on probing the robustness of these states against scattering, which is one of the key properties towards the application of topological devices.

Here, we present the first local probe STM measurements of MnBi$_4$Te$_7$ ($n = 1$) in both the real and momentum spaces. Combined with the observed ARPES band dispersion, we unambiguously assign the observed electronic band structures to the MnBi$_2$Te$_4$ septuple layer (SL) and Bi$_2$Te$_3$ quintuple layer (QL) terminations. Together with the input of ARPES (both regular and spin-resolved) and theoretical calculations, further investigation on the scattering process by quasi-particle interference (QPI) patterns unveils the spin configuration of their topological surface states. Unlike the surface state of a conventional TI (e.g., Bi$_2$Se$_3$) that is simply localized at the surface layer with attenuation into the bulk, the surface states of MnBi$_4$Te$_7$ shows strong hybridization between the magnetic layer and nonmagnetic layer at both terminations. For QL termination, a gap is formed due to the hybridization between the topological surface bands of the topmost QL and the bands from the neighboring SL layer, hiding the magnetism-induced gap near the Dirac point. For SL termination, our results suggest that a restoration of time-reversal symmetry occurs on SL termination, possibly due to the magnetically disordered surface. The spin texture of the topological surface bands shows a Rashba-splitting band from the QL underneath and a strongly warped band from SL, the hybridization of which contributes to the flower shaped QPI patterns. Our findings of such diverging topological behaviors on the two terminations of MnBi$_4$Te$_7$ strongly rely on the hybridization of bands from different building blocks of magnetic TI and non-magnetic TI, providing new insights in the surface-bulk correspondence of magnetic TIs and guidance to the heterostructure engineering in the emerging intrinsic magnetic topological systems.

## II. SURFACE TOPOGRAPHY

Fig. 1 illustrates the topographic images obtained by STM, where the two terminations are identified unambiguously. As shown in Fig. 1b, the crystal naturally cleaves at the QL termination of $Bi_2Te_3$ and the SL termination of $MnBi_2Te_4$ with step heights of 1.01 nm and 1.37 nm respectively. Comparing the zoom-in views of the QL (Fig. 1c) and SL (Fig. 1f) terminations, much more surface defects can be found on the SL termination. Even on the surface defect-free area, QL termination shows less corrugation, indicating lower bulk defect beneath the topmost layer (Fig. S2). Two types of major defects are found on the SL surface, categorized as bright and dark spots, the former of which is almost absent on the QL surface. To determine the origin of the defects, we obtain the atomic resolution images of the Te-terminated QL (Fig. 1 d-e) and SL (Fig. 1 g-h) surfaces at different bias voltages. On both terminations, the dark defects merge from triangularly placed holes to one triangular dark spot when the bias voltage is shifted from 0.5 V to -1 V, while the Te atoms from the topmost layer remains intact. We thus determine that these are $Mn_{Bi}$ anitsite defects (Mn replacing Bi at the second atomic layer), as seen in previous STM studies on Mn-doped $Bi_2Te_3$[27]. The bright spots are found to be slightly larger atoms within the surface lattice, as shown in Fig. S3, possibly attributed to $Bi_{Te}$ antisite origin (Bi replacing some of the top layer Te)[28]. The density of the detected $Mn_{Bi}$ anitsite defects is about 3.0% on the QL termination and 2.1% on the SL termination. Note that this is only for the second atomic layer as STM is surface sensitive. As shown in Fig. S2, on top of the $Mn_{Bi}$ defects on both terminations, there is a slight rightward shift (~10 meV) of *dI/dV* spectra compared with the defect free area, indicating a local *p*-type doping from this antisite defect. The migration of magnetic atoms (Mn), illustrated in Fig. 1a, may induce complicated surface magnetic order/disorder on both terminations, especially for the QL layer, as stated in previous studies that such amount of magnetic dopant is sufficient enough to induce long range ferromagnetic order in magnetically doped topological insulators[25, 29, 30].

**III. ELECTRONIC STRUCTURES**

We now focus on the defect free regions to obtain our averaged *dI/dV* spectra by scanning tunneling spectroscopy (STS). The results, as shown in Fig. 2a, reveal a great suppression of density of states (DOS) from ~ -285 meV to -350 meV on the QL termination, differing from the V-shaped, Dirac-like suppression of DOS near ~ -285 meV on the SL termination. The SL surface holds a higher density of states above -285 meV, while on the other side (below ~ -400 meV), the QL surface hosts higher DOS. Although it is not straightforward to determine gaps or positions of bands solely from the STS spectra, we can compare with our ARPES data to gain better understanding of the states. Fig. 2c and 2d are two representative Laser-$\mu$-ARPES spectra obtained from the same batch of crystals, representing the electronic structure on the two surface terminations of MnBi$_4$Te$_7$. In Fig. 2c, the outer electron pocket at $\bar{\Gamma}$ is seen to have a relatively flat bottom, where an apparent gap-like suppression of spectral weight is observed between $E = -270$ to $-320$ meV, in agreement with the STS at the QL termination (marked by the purple and the red arrows in Fig. 2c and Fig. 2a). Much higher ARPES intensity is observed for the bands below $E = -400$ meV, which is also reproduced by the high DOS below -400 meV bias voltage in the STS. We thus conclude that Fig. 2c shows the band structure of the QL termination. Similar analysis also reveals that Fig. 2d depicts the bands of the SL termination, where a high DOS is observed from $E_F$ to $E = -0.3$ eV. In this energy region, the bottom of one of the conduction bands at $E = -150$ meV can be identified in the STS by a DOS peak (brown arrows in Fig. 2d and Fig. 2a). Moreover, the X-shaped, Dirac-like linear dispersion and the gapless crossing point at $E = -285$ meV (red arrow in Fig. 2d) is endorsed by a V-shaped dip at the SL STS whose minimum locates at approximately -280 meV (red arrow in Fig. 2a). Therefore, by comparing with the STS spectra, we experimentally ascertain the assignment of the QL and SL terminations in the ARPES measurements for the first time.

We next perform first-principles calculations of $MnBi_4Te_7$ slabs to gain a better understanding of the electronic structures of the two terminations. For the QL termination, we assume that the local moment of the SL layers near the surface remains the same as that in the bulk, i.e., A-type AFM with out-of-plane spin orientation confirmed by neutron diffraction measurements [31]. The resulted band structure in Fig. 2e shows that the Dirac cone of the surface states is gapped by the magnetic proximity from the second SL layer, merging into the valence band at ~400 meV below the Fermi level (see the green box in Fig. 2e). More importantly, an indirect gap near the Dirac point energy occurs. The layer projection of the Bloch wavefunctions clearly shows that there is a band inversion between the conduction and valence bands, which is dominated by the topmost QL and the second SL layer, respectively. Therefore, the indirect gap is caused by the hybridization effect between these two layers. This is in sharp contrast to the conventional TI or magnetic-doped TI where the surface states around the Fermi level are governed by the surface layer, with attenuation into the bulk. It is also worth noting that other ARPES studies on the QL termination reveal a Λ-shaped intensity within the abovementioned hybridization gap[19, 20]. With the assumption of A-type AFM spin configuration, such an in-gap state can be interpreted as a hybridization effect between different orbitals contributed by the surface QL and its neighboring SL, respectively.

For the $MnBi_2Te_4$ SL termination (Fig. 2f), the experimental A-type AFM order with out-of-plane moment will inevitably open a sizeable gap of around 60 meV at the surface state (Fig. S4), inconsistent with our ARPES observation showing a gapless Dirac cone. In real materials, surface magnetism can be different from the bulk due to various reasons including surface reconstruction, disorder, or spin canting, etc.[32-34], which essentially governs the magnetic-related topological behaviors. Especially, considerable amount of $Mn_{Bi}$ anitsite defects are spotted (as discussed in Fig. 1) on both terminations, which may further complicate the surface magnetism. Considering some possible magnetic orders on the surface, we suggest a topmost nonmagnetic SL with disordered spin to be the most likely scenario, giving rise to a restoration of the time reversal symmetry on the SL termination. As shown in Fig. 2f, by assuming a nonmagnetic SL surface layer, we perform density functional theory (DFT) calculation and obtain almost linear dispersion with a negligibly gapped Dirac cone (gap size less than 10 meV), originating from the proximity effect of the third $MnBi_2Te_4$ layer, in good agreement with our ARPES results (Fig. 2d). Unlike the QL termination, the projection onto layers shows that the linear Dirac bands are mainly contributed by the surface SL layer. However, the strong hybridization between QL and SL layers also takes place as early as 150 meV below the Dirac point inferable from the contours of constant energy data shown in Fig. S9. Moreover, such hybridization can be captured by our quasiparticle interference patterns at higher energies (about 200 meV above the Dirac point).

## IV. SPIN-SELECTIVE QUASIPARTICLE INTERFERENCE

Having established the distinct band dispersion at the two terminations, now we carry out Fourier transform scanning tunneling spectroscopy (FT-STS) at various energies to further investigate the topological surface states. The disorder scattering will mix the states with different moment ($k_1$ and $k_2$) at the same contour of constant energy (CCE) for the surface bands, resulting in a standing wave with wavevector $(k_1 - k_2)/2$. Such interference patterns can be probed as local DOS modulation in the real space with wavelength $\lambda = 2\pi/q$, where $q = k_1 - k_2$ [35, 36]. Fig. S5-S7 displays the Fourier transform of spatial STS maps with a minimum linear dimension of 100 nm for both QL and SL terminations, evolving as a function of energy. In the simplest approach, the quasi-particle interference patterns should match the joint density of states (JDOS) as a function of momentum differences ($q$) between the two scattering states, $\text{JDOS}(q, E) = \int I(k, E) I(k + q, E) d^2 k$, where $I(k)$ can be experimentally determined from CCEs in the ARPES measurement. Fig. 3a and 3e shows typical CCEs of ARPES data for the QL and SL terminations at ~ 50 meV below the Fermi energy, respectively. Comparing with the calculated JDOS patterns from ARPES data (Fig. 3b and 3f), the FT-STS maps (Fig. 3d and 3h) display clear suppression of scattering intensity along $\bar{\Gamma} - \bar{K}$ directions. Previous FT-STS studies on topological insulators have suggested that the scattering process can be greatly affected by the spin texture of the topological surface states, as helicity only allows interference to occur when there is a finite spin overlap between the two scattering states[24, 25, 37, 38]. On the QL termination, a hexagonally warped shape of surface band structure is observed (Fig. 3a), accompanied by suppressed $\bar{\Gamma} - \bar{K}$ scattering in the QPI pattern. This resembles the warping-induced attenuation of $\bar{\Gamma} - \bar{K}$ scattering in $Bi_2Te_3$, where out-of-plane spin component can be introduced in between the warped corners, suggesting that the spin texture on QL termination can be understood in a similar manner.

Compared with the case of the QL termination, the QPI pattern of SL is more intricate. We first look into the results from spin-resolved ARPES for possible spin texture guidance at the SL termination. Fig. 2b represents a spin polarization ARPES map of $MnBi_4Te_7$ along the $\bar{\Gamma} - \bar{M}$ direction measured with a high-efficiency very low energy electron diffraction (VLEED) spin detector, showing the spin component along the tangential direction of the CCE band contours. The data is a mixed result of both the SL and QL terminations due to the millimeter-sized spatial resolution of the spin ARPES device. Interestingly, the electronic states of $MnBi_4Te_7$ shows an up-down-up-down spin configuration above the Dirac point energy (-280 < $E$ < 0 meV), resembling a typical Rashba-like spin splitting. We conclude such a spin pattern comes from the SL termination for the following reasons: (i) the *k*-space locations of the spin-polarized bands in Fig. 2b match those for the SL termination (Fig. 2d) to a better degree than those for the QL termination (Fig. 2c). Specifically, the spin pattern at moderate binding energies (i.e., $E_b$ ~ 0.2 eV) spreads to a wider range of $k_{\Gamma M}$ than the QL bands do (dashed curves in Figs. 2b and 2d). (ii) The inner ring in the QL termination (Fig. 3a) has been assigned as a bulk electronic state[13], where spin polarization is assumed to be far less polarized than that of surface states. Therefore, our spin-resolved ARPES results suggest a pair of Rashba splitting bands on the SL termination.

Taking the abovementioned experimental observations and first-principles calculations into account, we construct a uniform surface $k \cdot p$ model Hamiltonian to extract the central information of the electronic structures of both terminations, including Dirac bands, Rashba bands, hexagonal warping[39], hybridization between SL and QL layers and the exchange field due to the proximity effect from the bulk magnetism. It reads in the following

$$H_{\boldsymbol{k}} = \tau_+ H_{\boldsymbol{k}}^1 + \tau_- H_{\boldsymbol{k}}^2 + \Delta \tau_x. \qquad \text{Eq. (1)}$$

In Eq. (1), $H_{\boldsymbol{k}}^1 = \varepsilon_{\boldsymbol{k}}^D + v_F^D (\boldsymbol{\sigma} \times \boldsymbol{k})_z + \frac{\lambda}{2}(k_+^3 + k_-^3)\sigma_z$ describes a gapless, warped Dirac cone representing both the topmost $Bi_2Te_3$ layer for the QL termination and the topmost $MnBi_2Te_4$ layer with disordered local moment for the SL termination. $H_{\boldsymbol{k}}^2 = \varepsilon_{\boldsymbol{k}}^R + v_F^R (\boldsymbol{\sigma} \times \boldsymbol{k})_z + m_z^R \sigma_z$ describes a pair of Rashba bands with possible proximity-induced exchange field $m_z^R$ capturing the dispersion from the second topmost layers for both terminations. $\varepsilon_{\boldsymbol{k}}^{D/R}$, $v_F^{D/R}$ denote the parabolic dispersion and Fermi velocity of Dirac and Rashba bands, respectively, with $\lambda$ and $\Delta$ the parameters of hexagonal warping and hybridization strength between the top two layers, while $\tau$ and $\sigma$ are Pauli matrices with $\tau_\pm = (1 \pm \tau_z)/2$ and $k_\pm = k_x \pm ik_y$. The corresponding surface band dispersions and spin textures of both terminations, calculated from Eq. (1), is shown in Supplementary Fig. S12. We find that a direct gap is reproduced in the QL termination due to the combined effect of the hybridization $\Delta$ and exchange field $m_z^R$ (Fig. S12a), while a gapless Dirac cone beneath the Rashba bands generally captures the feature of SL termination (Fig. S12c). We next use the obtained spin textures in Fig. S12 to calculate the spin-selective JDOS and compare with the observed QPI.

The spin configurations for the QL and SL terminations are indicated by the arrows in Fig. 4a and b, respectively. We found that (i) the QL termination carries a spin-helical Dirac state with hexagonal warping, which gives rise to the canted spin texture along *z* direction; (ii) on the SL termination, a Rashba-split surface state comprises a pair of concentric circular bands with antiparallel spin helicity, between which lies another strongly canted helical state whose spin being antiparallel to the outer ring. Note that for the SL termination, hybridization can be observed when the strongly canted flower shape band crosses the outer circular band, as shown in Fig. S12. The spin-selective joint density of states (JDOS) can be estimated by adding a spin scattering matrix to the JDOS calculation via $SSJDOS(q, E) = \int I(k, E)T(q, k, E)I(k + q, E)d^2k$, where $T(q, k, E) = |\langle S(k, E)|S(k + q, E)\rangle|^2$[24] describes the scattering matrix where only the states with overlapped spin can be scattered. By assigning the abovementioned spin configuration to the CCEs of ARPES, shown in Fig. S10, the spin-selective JDOS maps at $E$ = -50 meV (Fig. 3c and g) show remarkable similarity with the QPI patterns on both terminations, with decreased scattering intensity along $\bar{\Gamma} - \bar{K}$ directions. On the other hand, the flower-shaped QPI pattern at SL termination differs from the spin-selective JDOS map on the petal. This could be possibly due to rearranged spin texture when hybridization occurs between the outer circular band and the strongly warped band, which unfortunately cannot be captured in our spin-selective JDOS calculation. Scattering from other energy cut may also play a role. However, despite the slight difference, the strong correlation between the QPI and spin-selective JDOS confirms the topological nature of the surface bands as well as the spin configurations on the two distinct terminations of $MnBi_4Te_7$, highlighting the hybridization between the strongly canted helical state from the surface SL layer and the Rashba-split states from its neighboring QL layer.

Consequently, as illustrated by Fig. 4a and b, the major scattering channel forming the edge of the petals in the QPI pattern can only be originated from $q_{SL}$ along $\bar{\Gamma} - \bar{M}$ with aligned spins for the SL termination, while for QL termination, the majority of allowed scattering occurs for wavevector $q_{QL}$. Detailed process of finding $q_{SL}$ and $q_{QL}$ is described in Fig. S11. We then overlay the dispersion of the wavevectors $q_{QL}$ and $q_{SL}$ obtained in the QPI patterns with the energy dispersion expected in the spin-selective JDOS calculated from ARPES data (Fig. S8-9). The plot shows good agreement between QPI and ARPES below Fermi level for both terminations, reassuring the nature of their topological bands. The dispersion for both vectors remains at high energy regime up to 400 meV (about 685 meV above the Dirac point), showing robust surface states.

## V. CONCLUSION

To summarize, we explore the distinct topological surface states of the two terminations of $MnBi_4Te_7$ by ARPES and STM. Compared with theoretical calculations, a gapless Dirac electronic structure is observed on the magnetic SL termination, whereas the nonmagnetic QL termination is found to be gapped because of the hybridization between different orbitals of the neighboring TI building blocks, implying different surface magnetism on the two terminations. Despite the microscopic defects of the magnetic atoms, the scattering process on both terminations can be understood by the topological surface states and the newly discovered spin configurations. The robust topological surface states remain present up to an energy regime far above the Fermi energy on both terminations, showing great tolerance to disorders. Our results provide insight not only to the impact of surface magnetism on the topological surface states, but also the potential of hybridization effect in the heterostructure engineering in magnetic TI systems. Further surface sensitive magnetic measurements or quantum transport study may be able to directly probe the detailed surface magnetic ordering on the two terminations and the related exotic quantum states.

**Acknowledgements**

The research was supported by National Natural Science Foundation of China under project No. 11674150, 11504159, 11674149, 11874195 and 11574128, NSFC Guangdong (No. 2016A030313650), the Key-Area Research and Development Program of Guangdong Province (2019B010931001), Guangdong Innovative and Entrepreneurial Research Team Program (2016ZT06D348 and 2017ZT07C062), the Guangdong Provincial Key Laboratory of Computational Science and Material Design (Grant No. 2019B030301001), the Science, Technology, and Innovation Commission of Shenzhen Municipality (JCYJ20160613160524999 and JCYJ20150630145302240), the Shenzhen Key Laboratory (Grant No. ZDSYS20170303165926217), and Center for Computational Science and Engineering at SUSTech. M. H. Xie acknowledges the financial support from a Collaborative Research Fund (C7036-17W) from the Research Grant Council, Hong Kong Special Administrative Region. We thank Jianpeng Liu, Haizhou Lu and Junhao Lin for valuable discussions.

**APPENDIX: METHODS**

The single crystal was prepared by flux growth method and confirmed by first single crystal X-ray diffraction data and the subsequent magnetic and transport measurements. The crystal has an AFM ground state with $T_N \sim 12$ K and a critical field of the spin-flop transition of $B_c = 0.2$ T, as shown in Fig. S1. The STM measurements were performed on in situ cleaved surfaces of $MnBi_4Te_7$ using commercialized STM of Unisoku 1500, operating at ~4.5 K with a base pressure better than $1 \times 10^{-10}$ mbar. Tungsten tips were modified in-situ by scanning on copper when needed. The dI/dV spectroscopy was obtained by a standard lock-in technique with frequency 687.2 Hz and an ac modulation of 5 mV. The spatial resolution for the FT-STS maps was 0.17 nm, giving a q-vector resolution of 0.004 Å$^{-1}$ in momentum space. For the QL termination, the JDOS data was calculated by the self convolution of raw CCE from ARPES, and a scattering matrix taking into account of the spin texture derived from $k \cdot p$ model (Fig. S12) is used for the spin-selective JDOS. For the SL termination, an idealized CCE constructed based on the ARPES CCE is used for the JDOS and spin-selective JDOS calculation, because of the complexity of the spin configuration in the topological bands. Details can be found in supplementary information.

ARPES measurements were performed at the Hiroshima Synchrotron Radiation Center (HiSOR). The spin-integrated spectra were obtained with an offline ARPES setup under 6.3 eV laser light, while the spin-resolved ARPES map (Fig. 2d) was obtained at Beamline 9B of HiSOR, using a very-low energy electron diffraction (VLEED) spin detector and 18 eV incident light.

The surface band structures of both terminations are calculated by density functional theory. We use the projector-augmented wave (PAW) pseudopotentials[40] with the exchange-correlation of Perdew-Burke-Ernzerhof (PBE) form[41] and GGA+U [42] approach within the Dudarev scheme as implemented in the Vienna ab-initio Simulation Package (VASP)[43, 44]. The energy cutoff is chosen 1.5 times as large as the values recommended in relevant pseudopotentials. The U value is set to be 5 eV. The *k*-points-resolved value of BZ sampling is $0.02 \times 2\pi$ Å$^{-1}$. The total energy minimization is performed with a tolerance of 10$^{-6}$ eV. SOC is included self-consistently throughout the calculations. The surface band structures of QL and SL terminations are obtained from slab calculations with the thickness of 9 and 7 van der Waals layers, respectively.

# FIGURES

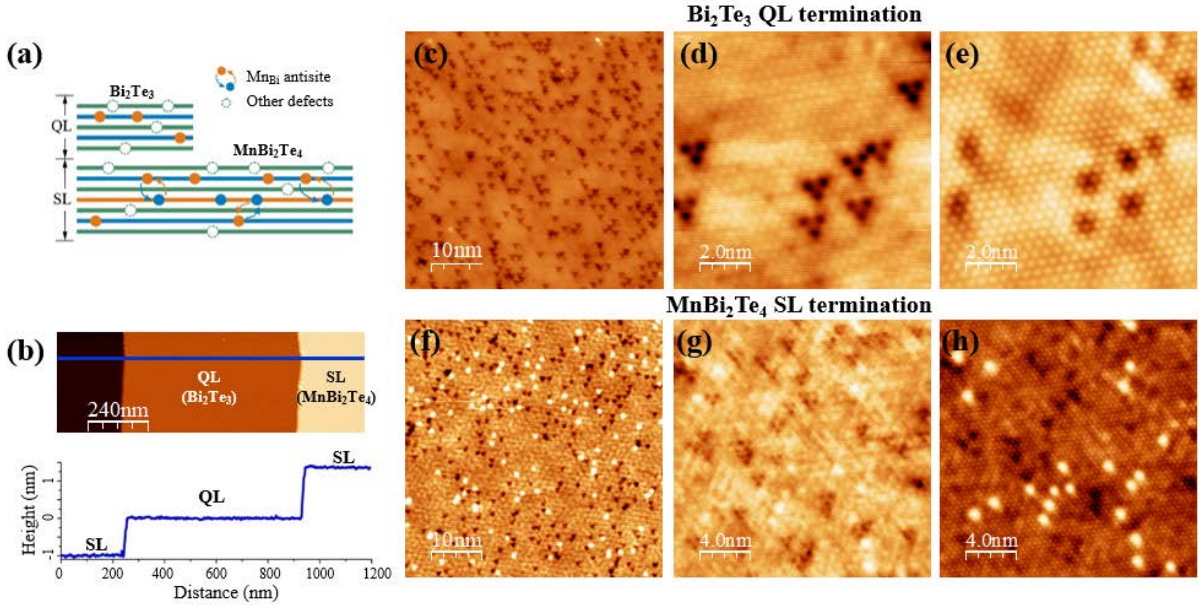

Figure 1. **STM topography images of the MnBi$_4$Te$_7$ surface.** (**a**) Schematic diagram of two terminations of the quintuple layer (QL) termination of Bi$_2$Te$_3$ and the septuple layer (SL) termination of MnBi$_2$Te$_4$ with atomic defects. (**b**) A large area topographic image of MnBi$_4$Te$_7$ surface showing two kinds of terrace steps. The lower part of (**b**) shows a line profile with step heights of 1.01 nm and 1.35 nm along the blue line, for the quintuple layer (QL) termination of Bi$_2$Te$_3$ and the septuple layer (SL) termination of MnBi$_2$Te$_4$, respectively. Considerable amount defects can be found in the zoom-in views of the QL (**c**) and SL (**f**) terminations. (**d-e**) and (**g-h**) are the atomic resolution images of the same areas at different biases on the QL and SL terminations, respectively, from which the majority of defect found in the QL layer is identified to be Mn-Bi antisite defects. For the SL layer, additional type of defect is found as bright dot at both positive and negative biases. Tunneling Parameters: (b) $V_{bias}$ = 1.2 V, $I_t$ = 20 pA (c) and (f) $V_{bias}$ = 1.0 V, $I_t$ = 20 pA (d) $V_{bias}$ = 0.5 V, $I_t$ = 1 nA (e) $V_{bias}$ = -1 V, $I_t$ = 1 nA (g) $V_{bias}$ = 0.5 V, $I_t$ = 200 pA (h) $V_{bias}$ = -1 V, $I_t$ = 200 pA.

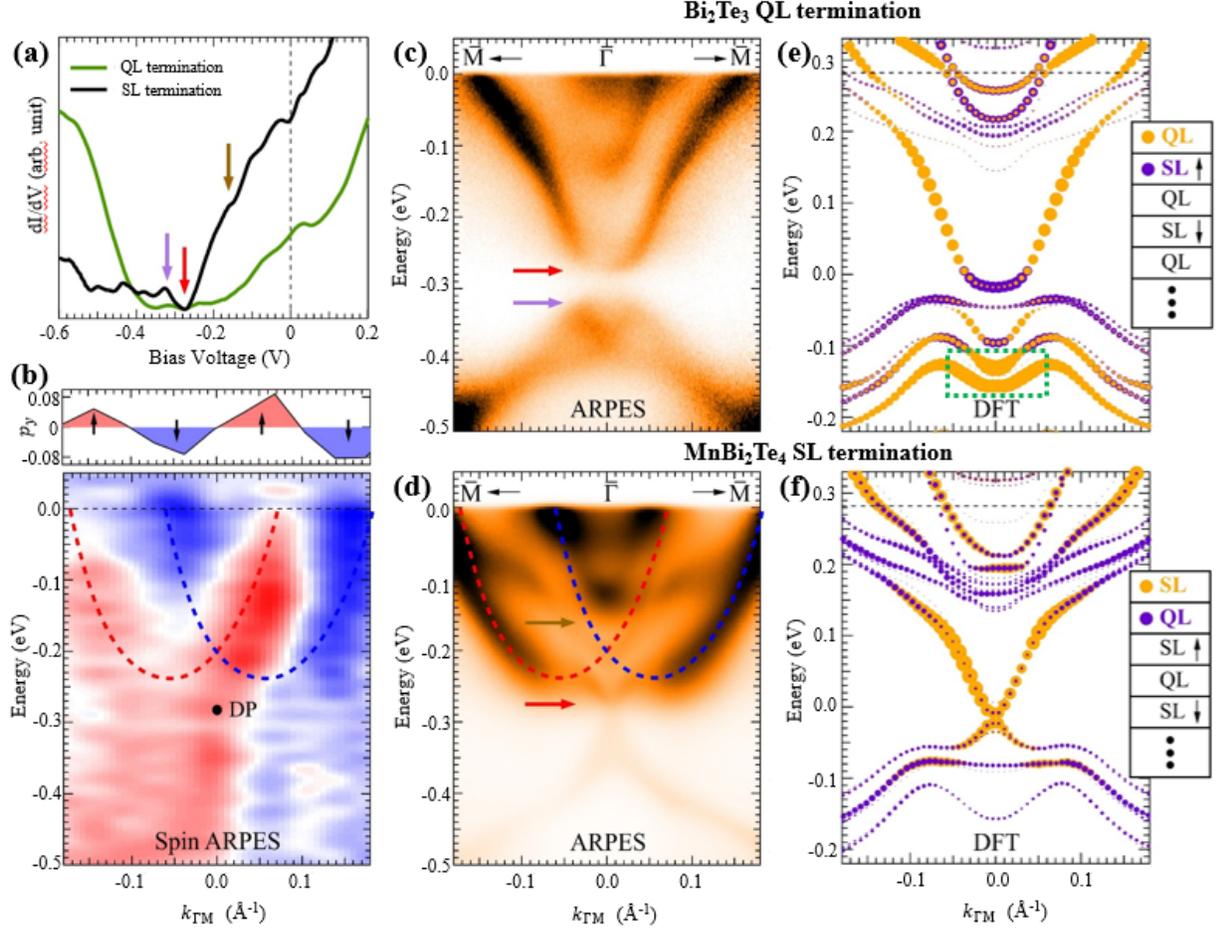

Figure 2. **Electronic structure of the two terminations of MnBi$_4$Te$_7$.** (**a**) Averaged d$I$/d$V$ spectra of a series of STS spectra obtained at defect free areas of the QL (green) and SL (black) terminations. Tunneling parameter: V$_{bias}$ = 0.5 V, I$_t$ = 500 pA. The arrows mark the hybridization gap opening on the QL termination (red and purple), the Dirac point (purple) and the onset of the inner Rashba splitting band on the SL termination (brown), in good agreement with the ARPES measurement. Dashed line marks the Fermi level. (**b**) Spin polarized ARPES map for both terminations of MnBi$_4$Te$_7$, with a line cut at $E_B$ = 0.08 eV showing the antiparallel helicity of the two bands. (**c**) and (**d**) are the ARPES intensity maps of the QL and SL terminations at the $\bar{\Gamma} - \bar{M}$ direction, respectively. Arrows mark the hybridization gap opening on the QL termination (red and purple) and the Dirac point (purple) and the onset of the inner band (blue) on the SL termination, highlighting important STS features in (**a**). Red and blue dashed curves in (**b**) and (**d**) are guides to the eye for the Rashba-split bands on the SL termination. (**e**) and (**f**) show the results from density functional theory calculations for the QL and SL terminations, respectively. The slab model assumes a non-magnetic SL termination for (**f**), while for (**e**), an A-type AFM spin configuration is used with the assumption that the magnetic order of the second SL is protected by the topmost QL. The green dashed box in (**e**) marks the gapped Dirac cone of the surface states.

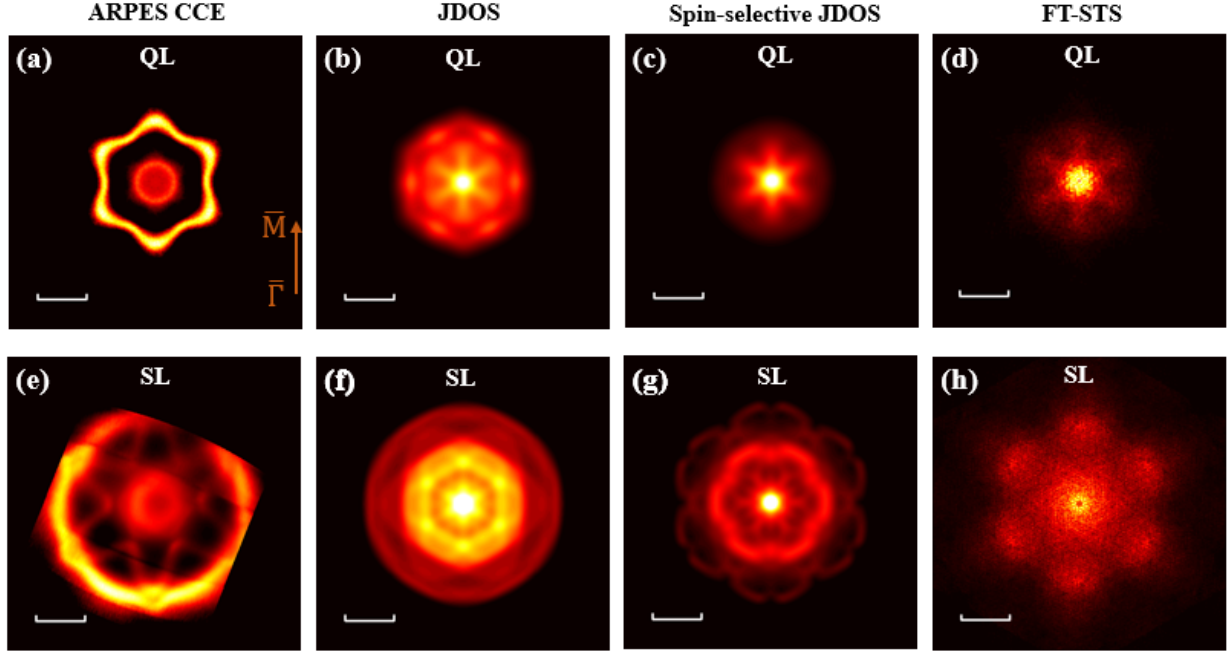

Figure 3. **ARPES Contours of constant energy, JDOS and spin-selective JDOS calculated from ARPES data, and FT-STS maps at -50meV of the two terminations.** Contours of constant energy (CCE) maps from ARPES measurements at 50 meV below Fermi level for QL (**a**) and SL (**e**) terminations. Brown arrow indicates the $\bar{\Gamma} - \bar{M}$ direction for all the maps. Joint density of states (JDOS) maps calculated from ARPESCCE maps are shown for QL (**b**) and SL (**f**) terminations. Intensity ratio of the bands is adjusted for the calculation on the SL termination. The JDOS maps show unexpected scattering intensity at $\bar{\Gamma} - \bar{K}$ directions. (**c**), (**g**) Spin-selective joint density of states for QL (**c**) and SL (**g**) terminations, by taking into account of the spin configuration described in Fig. 4. (**d**) and (**h**) are the Fourier transformation on the dI/dV maps (FT-STS) at $V_{bias}$= -50 mV, $I_t$ = 500 pA, image size 100×100 nm$^2$, showing good agreement with the spin-selective JDOS maps. The scale bars are 0.1 Å$^{-1}$ in CCE maps and 0.2 Å$^{-1}$ in the rest maps. Both ARPES CCE and FT-STS maps have been rotationally symmetrized.

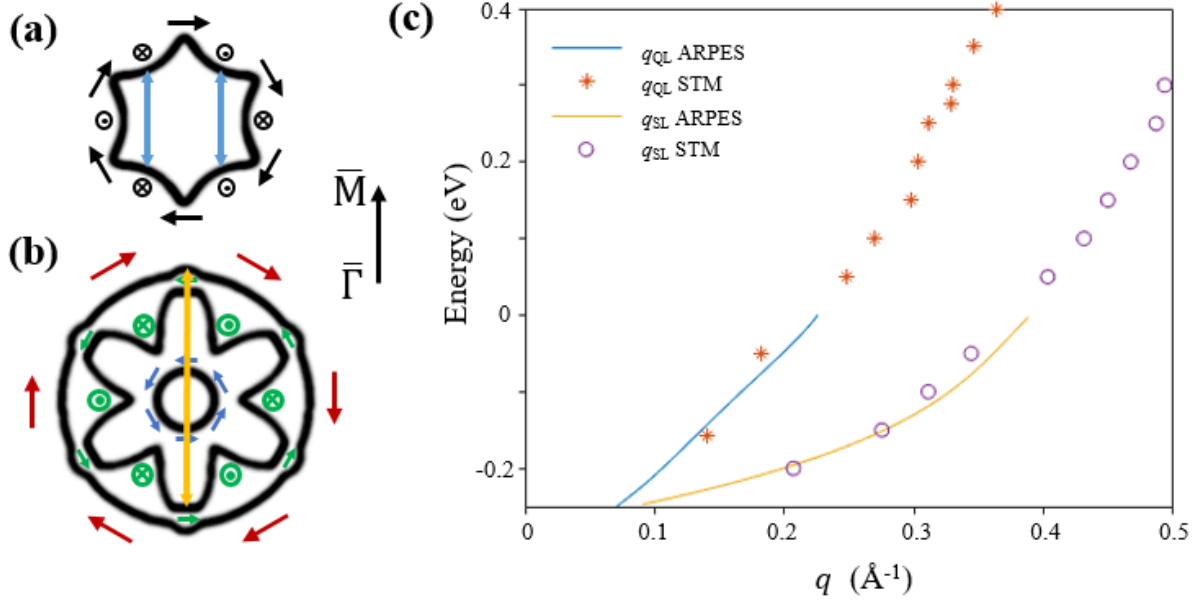

Figure 4. **Dispersion of $\bar{\Gamma} - \bar{M}$ scattering edges from ARPES and FT-STS.** (**a-b**) Schematic of CCEs resulting in the edges of $q$ vectors observed in the FT-STS maps along $\bar{\Gamma} - \bar{M}$ direction, arrows mark the spin orientation on these topological surface states: on the QL termination (**a**), the circular shape surface band is slightly warped for $E > $ -100 meV, causing canted spin along $z$, thus resulting in $q_{QL}$ marked by the blue arrows; on the SL termination (**b**), for $E > $ -150 meV, Rashba splitting contributes antiparallel helicity for the inner and outer rings in the CCE map (marked as blue and red arrows), where the outer ringhybridizes with a strongly warped surface band sandwiched in between. The middle flower-shaped band has the same helicity as the inner circular bands to fit the intensity along $\bar{\Gamma} - \bar{M}$ direction in the QPI map, leading to $q_{SL}$ marked by the yellow arrow. (**c**) Dispersion of the $q$ vectors calculated from the ARPES data (solid lines) on the QL (blue) and SL (orange) terminations, and the $q$ vectors obtained from the FT-STS data (orange stars for QL and purple circles for SL). The original CCE and FT-STS maps are shown in supplementary Figures S6, S7, S13 and S14.